# OBSERVATION OF DISCRETE OSCILLATIONS IN A MODEL-INDEPENDENT PLOT OF COSMOLOGICAL SCALE FACTOR VS. LOOKBACK TIME AND A SCALAR FIELD MODEL

H. I. Ringermacher* and L. R. Mead*
Dept. of Physics and Astronomy, U. of Southern Mississippi, Hattiesburg, MS 39406, USA

**Abstract**

We have observed damped longitudinal cosmological-scale oscillations in a unique model-independent plot of scale factor against lookback time for Type Ia supernovae data. We found several first-derivative relative maxima/minima spanning the range of reported transition-redshifts. These extrema comprise 2 full cycles with a period of approximately 0.15 Hubble times ( $H_0 = 68$ km/s/Mpc). This period corresponds to a fundamental frequency of approximately 7 cycles over the Hubble time. Transition-*z* values quoted in the literature generally fall near these minima and may explain the reported wide spread up to the predicted ΛCDM value of approximately *z* = 0.77. We also observe second and third harmonics of the fundamental. The scale factor data is analyzed several different ways including smoothing, Fourier transform and autocorrelation. We propose a cosmological scalar field harmonic oscillator model for the observation. On this time scale, for a quantum scalar field, the scalar field mass is extraordinarily small at $3\text{x}10^{-32}$ eV. Our scalar field density parameter precisely replaces the ΛCDM dark matter density parameter in the Friedmann equations, resulting in essentially identical data fits, and its present value matches the Planck value. Thus the wave is fundamentally a dark matter wave. We therefore posit that this scalar field manifests itself as the dark matter.

* E-mail: ringerha@gmail.com and www.ringermacher.com: Lawrence.mead@usm.edu

## 1. INTRODUCTION

In an earlier paper (Ringermacher & Mead, 2014) we developed an approach to plot scale factor against lookback time in a model-independent way for the first time and performed a preliminary analysis to locate the transition-time (or redshift) of the universe. We pointed out that transition-z values found in the literature span a wide range from *z* = 0.45 to *z* = 0.78. The ΛCDM prediction is approximately 0.77. In that paper we performed considerable smoothing of the data and found no absolute minimum out to at least *z* = 0.6 at which point the data was quite "noisy". We did, however, find apparent oscillations, but ignored and smoothed them out since one can in fact create oscillations by filtering wide-band noise, depending on the filter cut-off frequency. In the present work we analyze the apparent oscillations more carefully. We emphasize that the main purpose of this paper is to describe the oscillations. The model we develop to support our observations is secondary, simple and, at present, incomplete in that it describes only the dominant frequency. Its purpose is to demonstrate that such oscillations can propagate into the present and to provide a platform to estimate the mass of the scalar field by matching the oscillation frequency to the observation. It is convenient to work in frequency units when analyzing the oscillations. We shall define 1 cycle over 1 Hubble time as 1 Hubble-Hertz or 1 HHz. In the analysis as well as in the model time is normalized to the Hubble time.

For our analyses, we perform an FFT and autocorrelation in addition to multiple smoothing algorithms and statistical analyses on both the $a(t)$ data and its first derivative. An oscillation at a dominant frequency of approximately 7 cycles over one Hubble time, which we term 7 Hubble-Hertz (HHz) is revealed as are a second and third harmonic from the FFT. Since the oscillations are seen in the scale factor, one might expect they should also be seen in the distance modulus data. Analysis, based on our model, shows that the scale factor plot has a signal-to-noise ratio (SNR) advantage of a factor of 3-5 over the Hubble diagram for $z<1$. Also the signal is a complicated shape and not periodic when plotted against redshift. These facts would reduce the likelihood of an observation in the Hubble diagram. However, since our observed amplitudes were 2-3 times greater than our model, it may be possible to extract this signal with care from the modulus plot.

We also present a scalar field model that supports the presence of such discrete oscillations into the present epoch. Many authors have tried similar models to attempt to explain dark energy or dark matter (Ratra & Peebles, 1988; Khoury, et al., 2001; Gao, et al., 2010; Suárez, Robles & Matos, 2013 ) but had no experimental foundation upon which to build. We demonstrate that a simple damped harmonic oscillator scalar field coupled to the Friedmann equations describes the dominant oscillation and produces a near-perfect scale factor fit to ΛCDM.

This paper is organized as follows. The data analysis is first described. Smoothing algorithms are used on the scale factor vs. lookback time data set and its first derivative. Next, an FFT is performed on the same data. A simulation of the oscillations with the measured amplitude and damping along with a realistic noise distribution is carried out and a statistical estimate of the likelihood that the signal is real is performed using FFTs in blind testing. The scale factor data is then analyzed a final time in Section 4 with improved processing, then using autocorrelation and discrete Fourier analysis. This resulted in a higher SNR at the fundamental frequency. Finally the scalar field model is presented. We add that the model does more than describe the oscillations – the matter density deriving from the scalar field is precisely equivalent to the ΛCDM dark matter density in the Friedmann equations by substituting for it and encompasses the same perfect fit to the SNe data. It successfully substitutes for dark matter over the measured range of SNe. This will be detailed in the Model section of the paper.

## 2. SCALE FACTOR DATA SMOOTHING ANALYSIS

### 2.1 Differences between the Hubble diagram and a model-independent scale factor plot

We employ the same data set as was used in a previous paper ( Ringermacher & Mead, 2014 ). This is a combination of SNe data of Conley et al.(2011) and Riess et al.

(2004) together with radio galaxy data of Daly & Djorgovski (2004) out to $z = 1.8$. The $a(t)$ plot is obtained by initially calculating $a(t)$ directly from the redshift and evaluating the corresponding lookback time as described in the paper. Lookback time is then sorted and $a(t)$ is plotted for each time. This effectively transfers any lookback time noise into $a(t)$. Although the entire set was analyzed, most of the oscillation was observed in the Conley data alone and could be analyzed as such. We note that our scale factor plot is a function of time and thus an analysis can reveal discrete frequencies. The standard Hubble diagram of distance modulus against redshift would present oscillations as a *z*-dependent complicated spectrum. Our data analysis simulation indicates that the scale factor plot analysis also has an SNR advantage of as much as a factor of three to five for $z < 1$ compared to that found directly from a Hubble diagram. The advantages subside for $z > 1$. Thus with the availability of more high-*z* SNe and knowing what to look for, it may be possible to see the oscillation directly from the Hubble diagram. We describe the expected signal shape and amplitude in the model discussion.

### 2.2 Raw data smoothing

Two types of smoothing were used – median smoothing and Gaussian smoothing. Since the $a(t)$ data is slowly varying with time, a derivative of the smoothed, unbinned, data was first attempted to flatten the curve permitting later binning of the data. This would reveal any inflection points as a minimum in the derivative. Median smoothing was chosen to process the raw, unbinned, $a(t)$ data prior to differentiation. There are 527 unequally spaced points in the data set. The point density is very high at low *z* and sparse at high *z*. We used a running 49-point window resulting in a filter which, over the region of oscillation, has an effective 7 HHz, 3 dB cutoff. We also tried a running 17-point window (15 HHz cutoff) and found essentially the same final results. The 7 HHz signal was not caused by filtering effects or by binning effects as was vindicated by our simulation and more advanced analysis below. It also was not caused by joining data sets since most of the signal was observed in the Conley set alone. After smoothing we differentiated the data. The choice of smoothing and differentiation techniques is dominated by the need to extract a signal whose width is on the order of 10% or less of the time base. Thus we do not use the very wide (20-30 %) time window smoothing algorithms sometimes seen in redshift data analysis.

### 2.3 Smoothed data derivative

A simple centered 3-point (2 time bins) derivative would introduce a great deal of noise. Instead we chose a wide-baseline 3-point derivative sampling a 10-bin span or greater. This type of differentiation is the numerical analog of magnetic field-modulation typically used in Nuclear Magnetic Resonance wide-line spectral analysis (Bozarth and Chapin, 1942). It is capable of enhancing broad, weak signals when the modulation width (time window width here) is equal to the line-width (oscillation period here). The signal must be “line-like” (here a damped oscillation). For example, a 10-bin derivative will improve the SNR by a factor of approximately 5 over a 2-bin derivative. This is further described in Appendix A.

Figure 1 shows the plot of unequally spaced scale factor vs. lookback time data, the smoothed data, a 30-bin derivative of the smoothed data and the final, Gaussian-smoothed oscillations. The final Gaussian smoothing uses a time window of 0.08 (8% of the normalized time scale). Our simulation shows that this series of filters has an effective cutoff in the neighborhood of 7 HHz. The frequency measured here, based on an average 2-cycle period of 0.155 Hubble times, is 6.5 ± 0.6 HHz. A best match of the oscillations in Fig.1 to the scalar model was found at a 6.95 HHz frequency. Early time points between 0.3 and 0.37 were distorted by the smoothing and were discarded. End point data very near $z = 0$ was also affected by the smoothing and is not reliable. The relative minima were found at normalized lookback times $t = 0.78$, 0.63 and 0.47. The relative maxima were found at normalized lookback times $t = 0.87$, 0.71 and 0.56. In turn, these times correspond respectively to red-shifts of $z = 0.26$, 0.51, and 0.9 for the minima and $z = 0.14$, 0.37, and 0.66 for the maxima. In addition to the four processed sets we display the time derivative, $\dot{a}(t)$, of the ΛCDM scale factor (rising curve) for WMAP Omegas. We note that the oscillations are centered directly on this curve. We have multiplied the derivative by a factor of 0.2 to enable a common display on the plot hence the slope 0.2 as $t$ approaches 1 rather than slope 1 as seen for the scale factor. Here the signal amplitude after smoothing is approximately equal to the RMS noise, 0.1. Section 4.3 will describe an improved analysis.

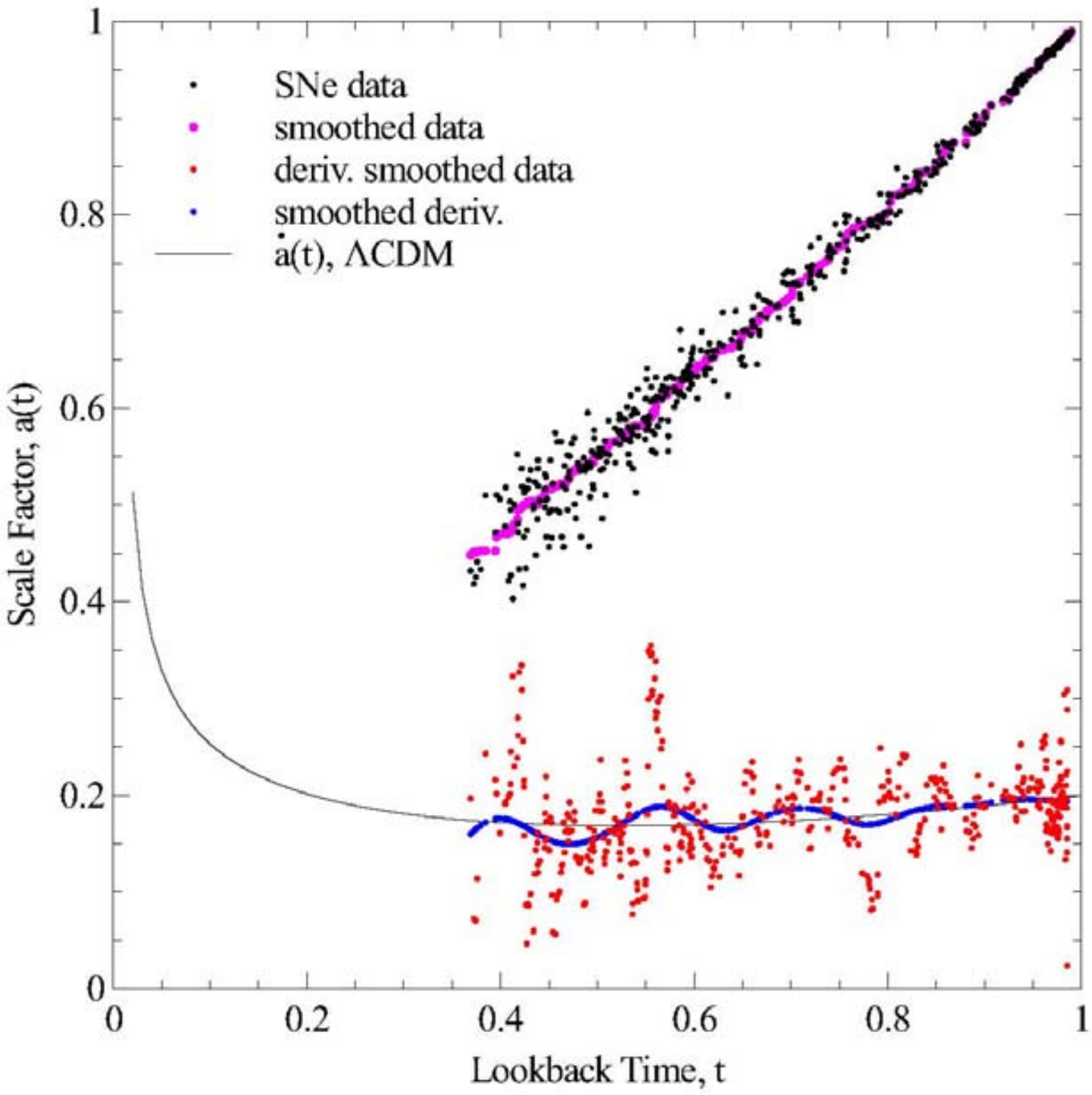

**Fig. 1.** Plot of scale factor SNe data, smoothed data , derivative of smoothed data, smoothed derivative and $\dot{a}(t)$ for ΛCDM Omegas.

### 2.4 Comment on unequally spaced redshift data

Since the redshift data are unequally spaced, but putatively random, so too is the scale factor data. Fourier transform analysis requires equally spaced data, so for that analysis the data were binned. We have shown that our redshift data can be divided into three groups of spacings; nearby tightly spaced redshifts $z < 0.04$, midrange $0.04 < z < 1$, and high sparse redshifts $z > 1$. The midrange grouping comprises 360 points with a mean spacing of $\Delta z = 0.0028$ and standard deviation $\sigma = 0.003$. How random is this? We can ascertain that by generating an artificial set of 360 randomly spaced points between redshifts 0.2 -1 and examining those statistics. We find for that set, $\Delta z = 0.0028$ and $\sigma = 0.003$. Thus we can say with confidence that the midrange real redshift data set appears to be randomly spaced. The unequally spaced data points can be plotted against their equally-spaced point indices. This produces a line of constant average slope with effective random noise. That is the noise that would be introduced into an analysis otherwise requiring equal data spacing.

## 3. SCALE FACTOR FFT DATA ANALYSIS

An FFT analysis was performed on the data to confirm the presence of discrete frequencies. This was done in several different ways. In all cases the data required equal-time binning. We chose either 256 or 512 equally spaced time bins. Any unfilled bins were randomly extrapolated between bounds. We performed FFTs on binned raw $a(t)$ data, smoothed-then binned $a(t)$ data and the first derivative of the data. We describe the results below.

### 3.1 FFT of 256-binned data after smoothing

The raw $a(t)$ data of Figure 1 was first smoothed as is with unequal times. 49-point median smoothing was performed on the $a(t)$ curve as before. It was determined in the simulation that the 49-point smoothing performed on the 527 unevenly spaced points was equivalent to a 27 point smoothing over the region of interest in the 256-binned set. The data was then sorted into 256 bins from $t$ = 0.3 (the earliest time data) to $t$ = 1. This necessitated renormalizing the FFT frequency scale. A third order polynomial was fitted to the smoothed data and subtracted from it to flatten the curve, removing any DC component, and permitting a clean analysis of the residuals. Any noisy data prior to $t$ = 0.4 was removed and the flattened curve zero-padded to $t$ = 0.3. Inclusion of excessively noisy data can overwhelm the signals. An FFT was then performed on 256 points between $t$ = 0.3 and $t$ = 1. Results are shown in Figures 2 and 3.

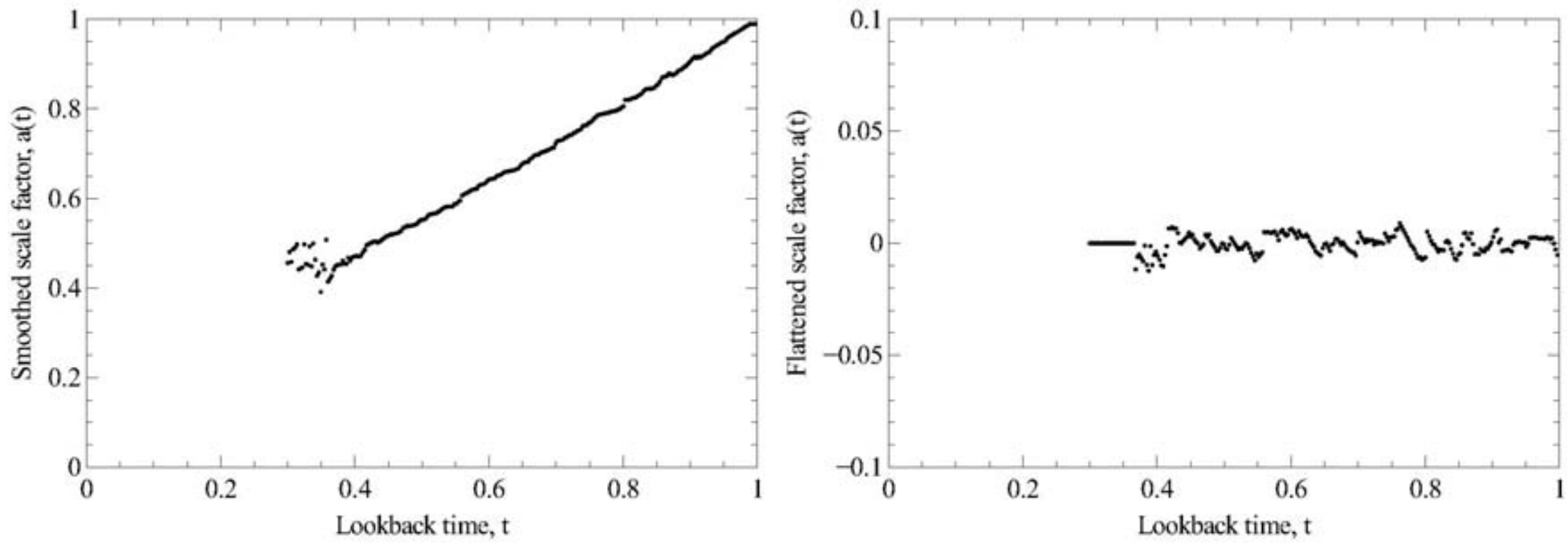

**Fig. 2.** Plot of 49 point smoothed scale factor data (left) with 256 equal-time bins from $t = 0.3$ to $t = 1$. Noisy data is seen at early times. Plot of flattened and padded scale factor data (right) with 256 equal-time bins from $t = 0.3$ to $t = 1$. The noisy data has been discarded.

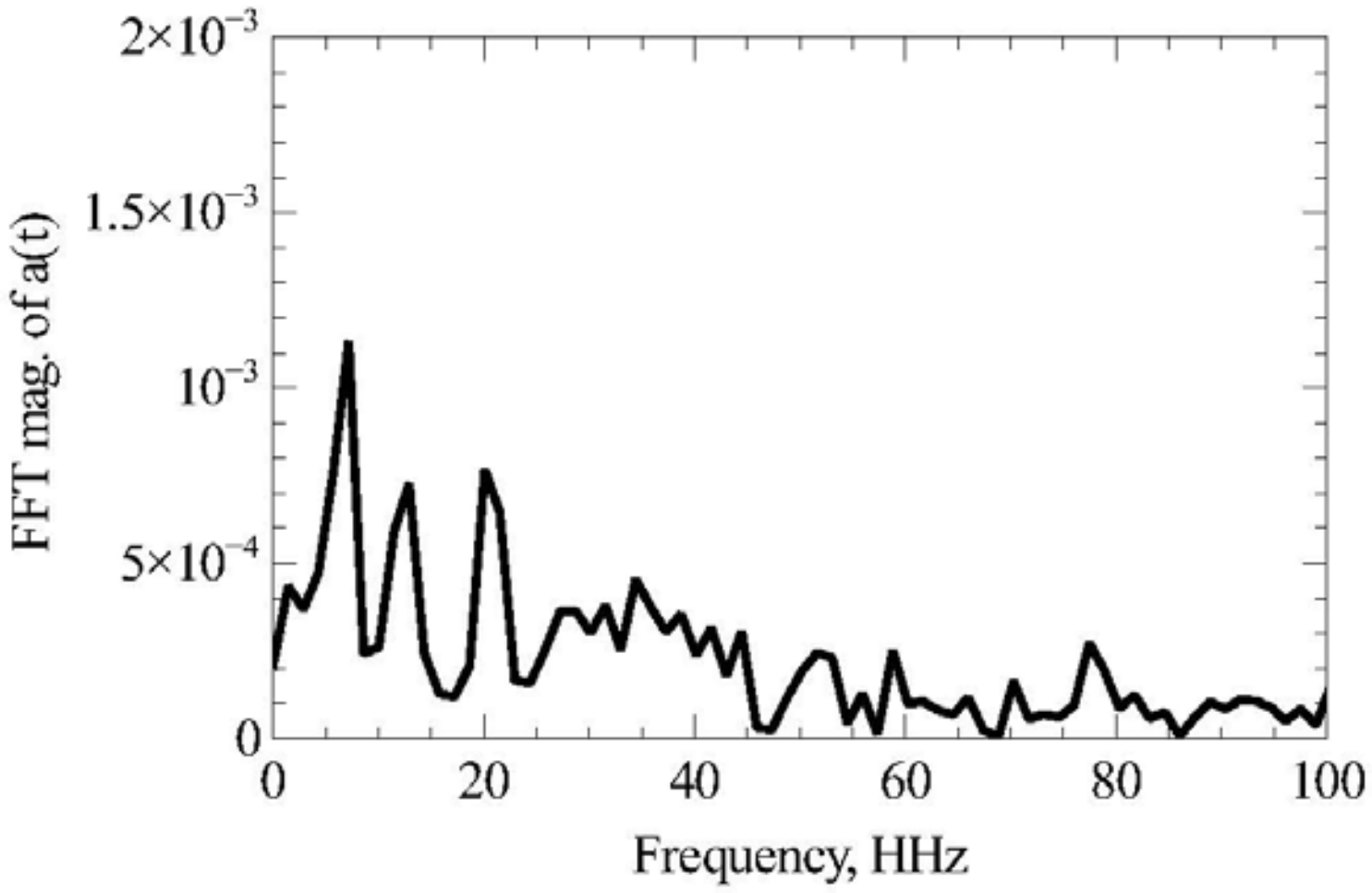

**Fig. 3.** FFT of Figure 2 data with frequency scale in units of HHz. Frequencies at 6.5, 13 and 20 HHz are prominent.

The FFT confirms the dominant oscillation between 6.5 – 7.0 HHz and finds, in addition, second and third harmonics. The harmonics would be reduced in amplitude since the raw unequal time data was smoothed with a 49 point median filter – equivalent on average to having approximately a 6 HHz, 3 dB, cutoff. These harmonics appear in nearly all our analyses, so we report them. The fact that these signals appear in the equal-time binned data means the unequally spaced redshifts were not their source.

### 3.2 FFT of 512-binned data after smoothing

We repeated the 256-bin procedure above but binned the raw, smoothed, data to 512 points instead to obtain the FFT shown in Fig.4.

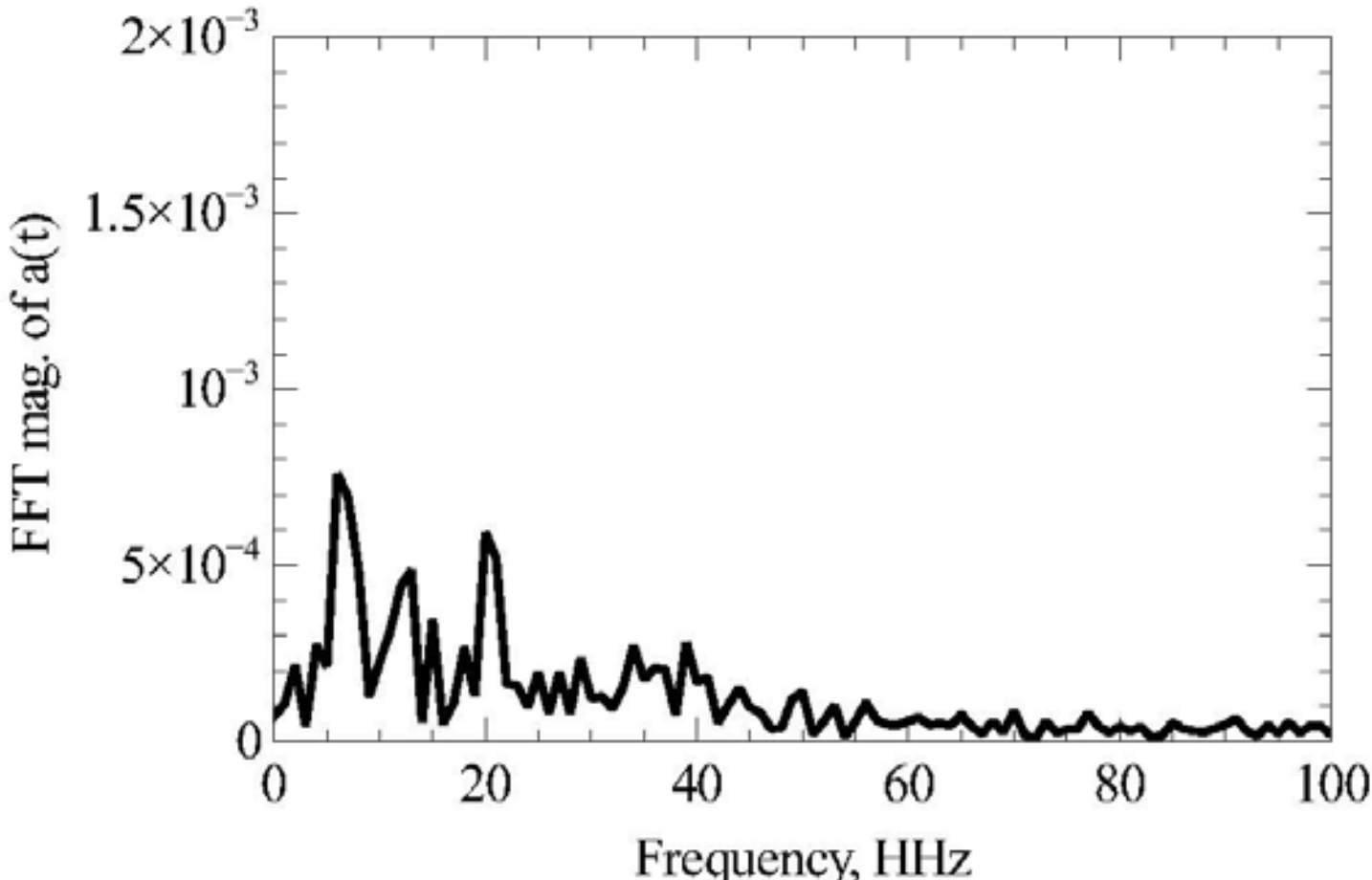

**Fig. 4.** FFT of 512-binned, smoothed, data with properly normalized frequency scale in units of HHz. Frequencies at 6.5, 13 and 20 HHz are prominent.

### 3.3 FFT of smoothed data after binning

We next tried first binning the unsmoothed flattened raw data into 256 equally spaced time bins (Figure 5, left), followed by smoothing (Figure 5, right). This had a significantly different noise spectrum since the binning itself smoothes the data – in ways that can sometimes be undesirable. This occurs because the sparse data is binned in time slots exceeding the quoted experimental error, thus potentially losing information. The smoothing window is 0.02 with a 3 dB cutoff at 18 HHz. Figure 6 shows the FFT of the data of Figure 5. Frequencies at 6.5, 13 and 19 HHz are detected. The signal at 3 HHz is likely due to imperfect flattening.

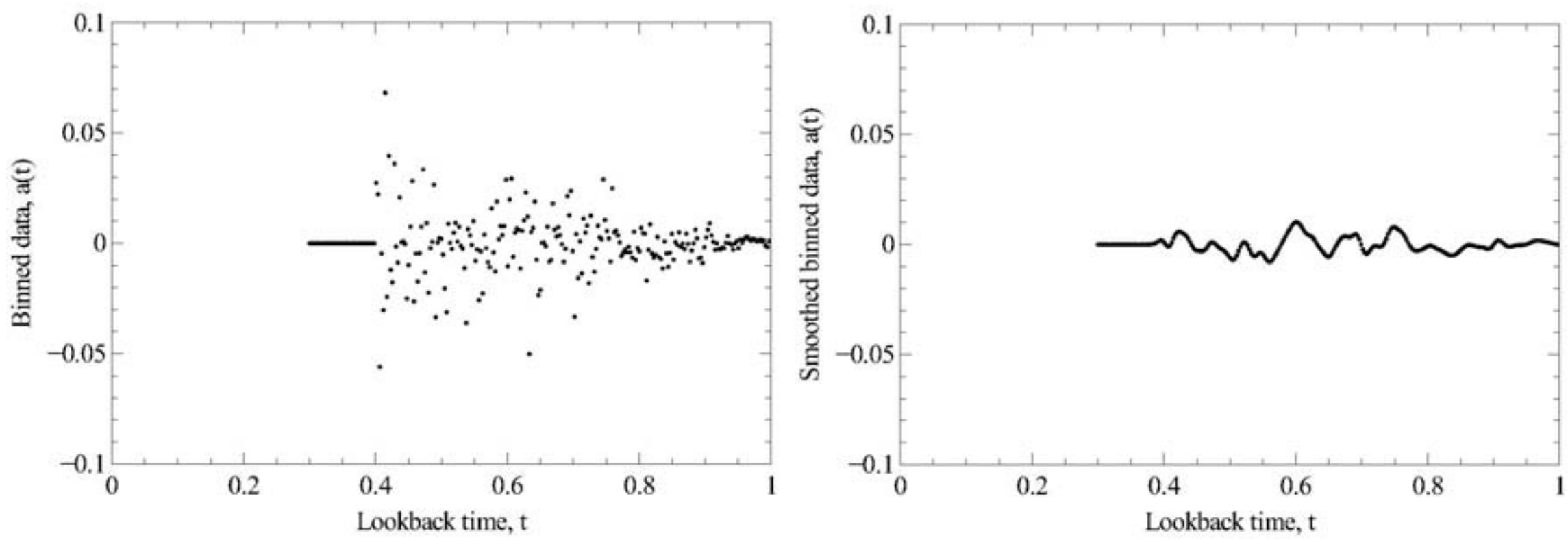

**Fig. 5.** Plot of flattened, padded and binned raw data (left). Gaussian-smoothed data(right). Smoothing window was 0.02, with an 18 HHz , 3dB cutoff.

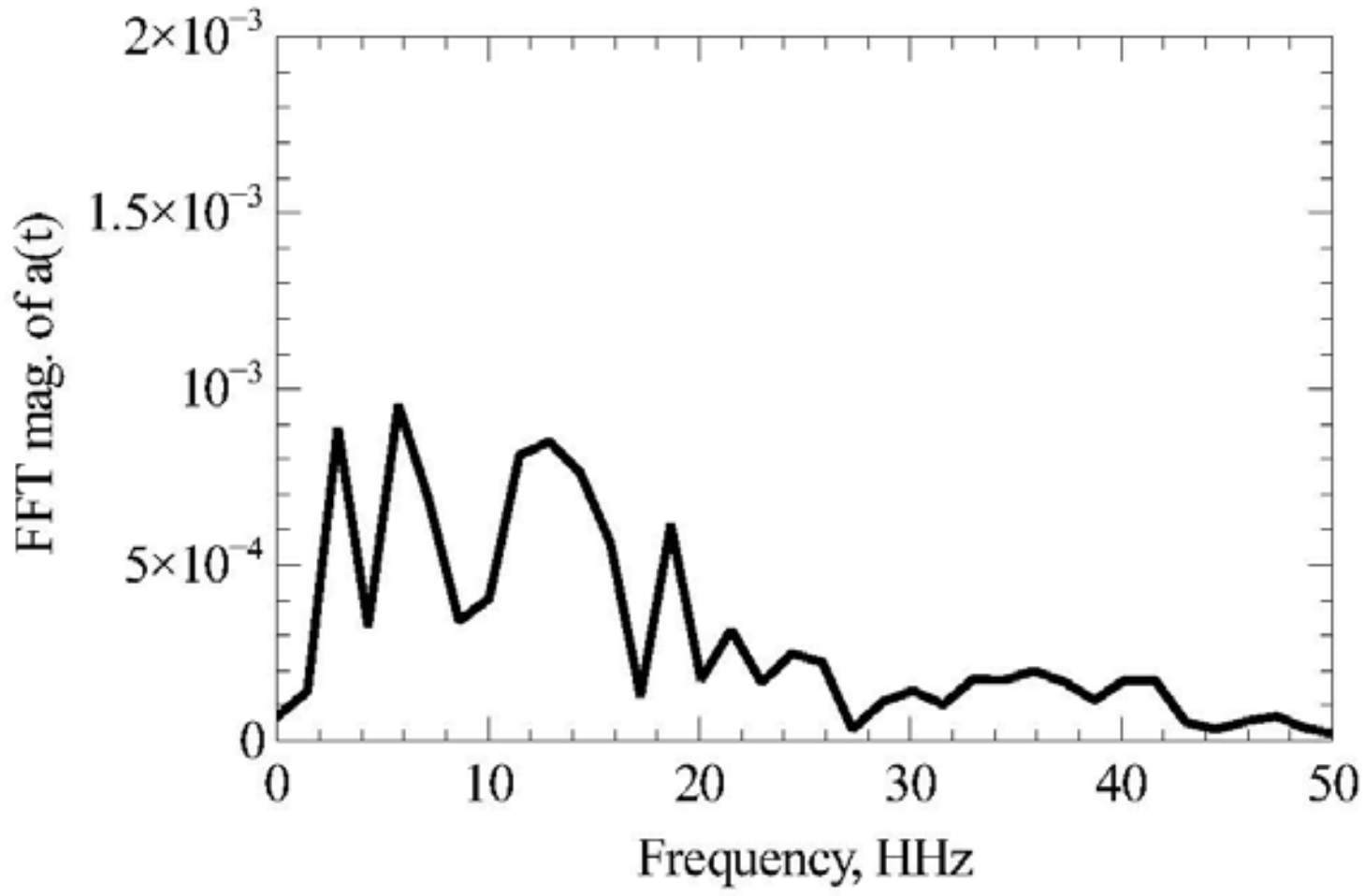

**Fig. 6.** FFT of Fig. 5. 6.5, 13 and 19 HHz signals are seen. In addition a 3 HHz signal is seen and may arise from imperfect flattening of the data after binning.

### 3.4 FFT of derivative of smoothed, binned data

Figure 7 shows the FFT of the derivative of the data of Figure 2. The data was not first flattened. Consequently there was a residual DC component in the FFT. Otherwise the same frequencies as before remain. Recall, this data was smoothed before binning. A 3-point derivative over a running 10-bin baseline was then used. The narrow derivative window emphasizes the 20 HHz signal and improves the SNR.

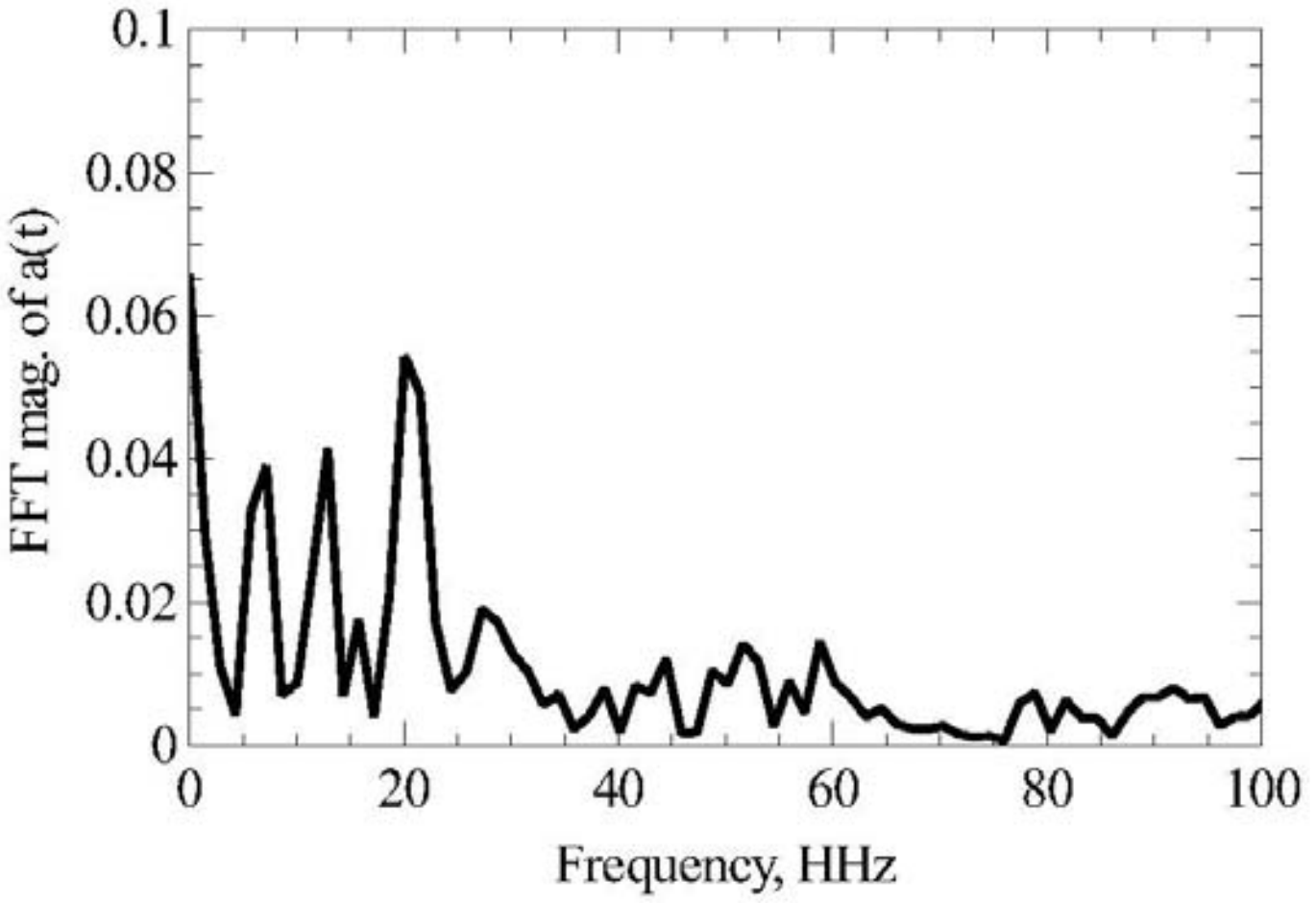

**Fig. 7.** FFT of derivative of Fig.2. 7, 13 and 20 HHz signals are seen.

## 4. EXPANDED ANALYSIS

We further analyze the signal by building a signal synthesizer with a waveform simulating our scalar field model presented in Section 5 and examining the signal with noise statistics. This section is concluded with an improved analysis approach to maximize the 7 HHz signal. This is followed by an autocorrelation analysis that confirms the presence of the signal at higher SNR.

### 4.1 Model waveform

We create a model of the waveform based on our scalar field model for further testing. This model is good in the range $t = 0.2 - 1.0$. We include, at this time, only the fundamental frequency at approximately $f_0 = 7\,HHz$, damped exponentially to match the decay rate in our model and amplitude-adjusted to match the observations at $t = 0.6$. In our model, the oscillatory component rides effectively on the ΛCDM solution to the Friedmann equations. The signal used to model our observation in Figure 8 (right) is intended to be the "residual" signal after the ΛCDM $a(t)$ has been removed, plotted against lookback time and has the form

$$a(t)_{residual} = 0.028\, sin(2\pi f_0 t(z))\, e^{-2.8\,t(z)}, \tag{1}$$

where,

$$t(z) = 1 - \int_0^z \frac{dx}{(1+x)\sqrt{0.27(1+x)^3 + 0.73}} \quad . \tag{2}$$

is the model-based lookback time. When this signal is placed onto a z scale, corresponding to the lookback time, the result is the $z$-model curve of Figure 9 plotted against $z$ out to $z = 2$. This is what the waveform residual would look like on a plot of distance modulus against redshift with the approximate (observed) amplitude.

### 4.2 Statistical tests from simulations

The likelihood of seeing a 7 HHz signal from pure noise was simulated. The data noise scatter from $t$ = 0.3 to $t$ = 1 was approximated as a linearly rising Gaussian normal distribution about a zero mean with 0.07 variance. This distribution simulated the flattened data set scatter very closely. The signal used was derived and scaled from the solution of $a(t)$ from the scalar field model discussed in Section 5. Figure 8 compares the true and simulated noise scatter of the residuals on a plot of $a(t)$ vs. $t$. Also shown is the 7 HHz scalar model wave signal used for trials. The amplitudes are correct. The amplitude of the 7 HHz signal used for the trials is about twice that for the model solution – closer to what was observed. The waveform is described in more detail in Section 4.

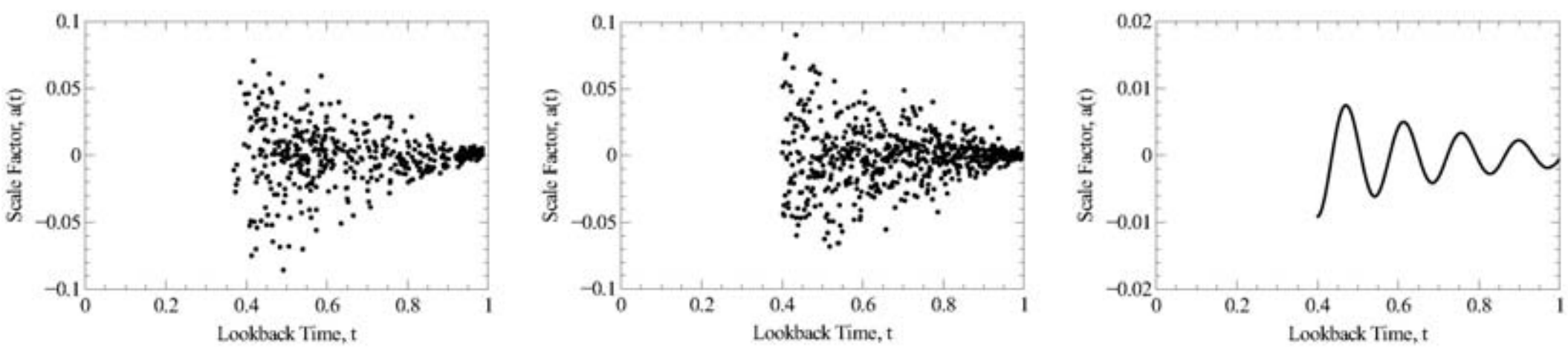

**Fig. 8.** Real noise scatter of flattened SNe data is left. Simulated scatter is center. Simulated signal is right.

The peak-to-peak noise scatter in the region of interest around $t = 0.6$ was approximately 0.08 on a scale of 0 – 1. Gaussian smoothing with a window of 0.02 (3 dB, 18HHz cutoff) was used on noisy data looking like $a(t) = t$, close to reality. First, only noise was introduced with zero signal. 500 trials were run. A 7 HHz signal appeared approximately 5% of the time. Next a 7 HHz signal was introduced at 1/10 the noise peak amplitude. 500 trials were run. A 7 HHz signal level at least twice the noise was seen approximately 50% of the time. Thus the likelihood of the dominant signal at 7 HHz being real is 10/1 using this filtering. No simulation we could compose produced harmonics as seen in the real data, including padding trials. The harmonics were generally lower amplitude except those of Figure 7 where a 10-bin derivative was used which emphasizes the higher frequencies.

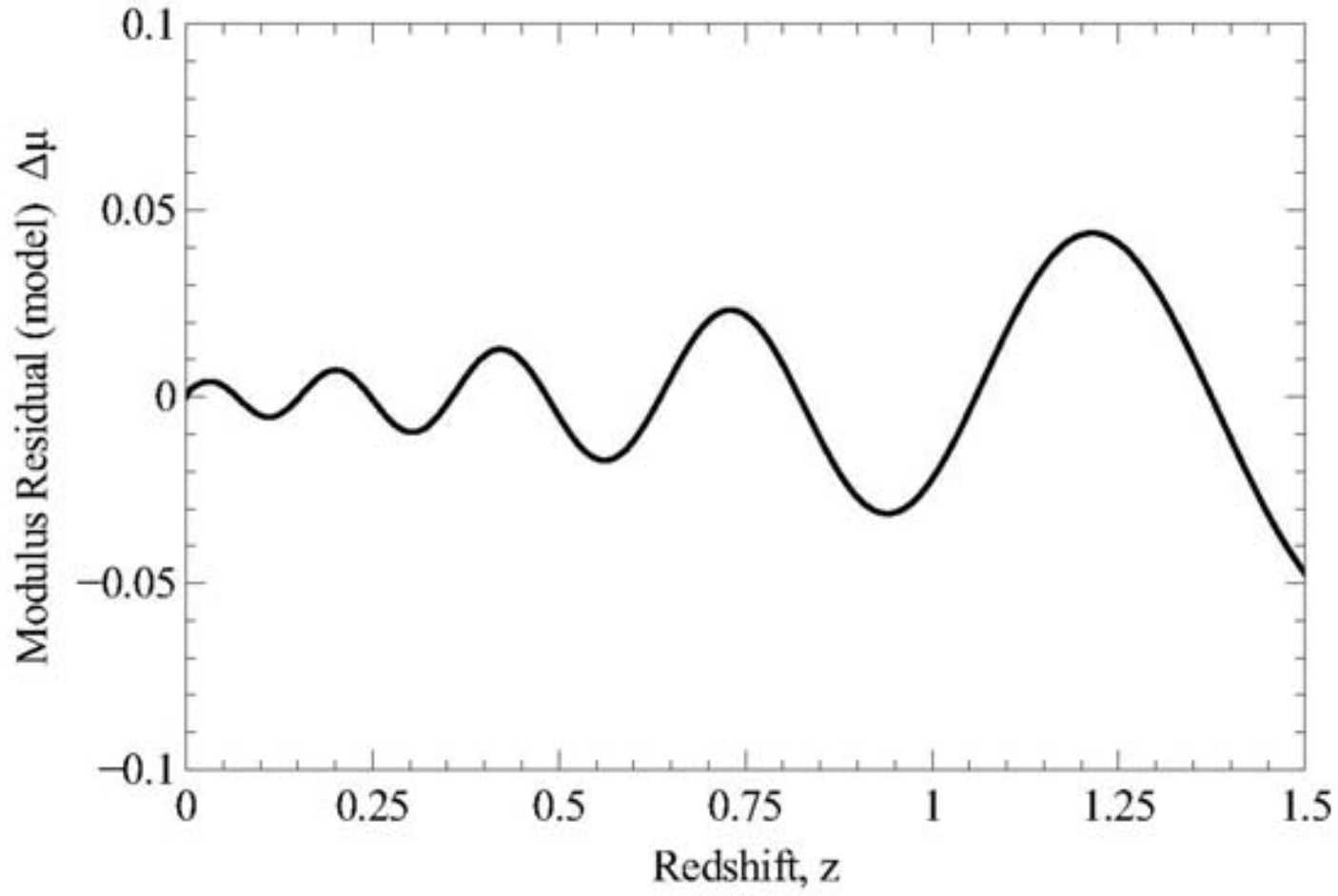

**Fig. 9.** Residual oscillation in distance modulus vs. z at 7 HHz to z = 1.5.

### 4.3 Scale factor autocorrelation analysis

We analyze the "flattened" raw, unbinned, data of Figure 8 (left) using Gaussian smoothing. The unbinned data will introduce some random noise as described in Section 2.4 but may show stronger results. The earlier analyses smoothed the $a(t)$ data before flattening. We use a 0.03 smoothing time window followed by a 30-bin derivative on the smoothed data to maximize the 7 HHz component. Once again, only the data for $t \geq 0.4$ (527 points) is analyzed to exclude early-time sparse-data noise. We use the standard definition of the auto-correlation index, $r_k$, where $k : \{1, N-1\}$ and $\mu$ is the mean of the set of $N$ values:

$$r_k = \frac{\sum_{i=1}^{N-k} (\dot{a}_i - \mu)(\dot{a}_{i+k} - \mu)}{\left( \sum_{i=1}^{N-k} (\dot{a}_i - \mu) \right)^2} \tag{3}$$

Figure 10 shows the results of the autocorrelation and discrete Fourier transform performed on this data. The Fourier transform (Fig. 10, right) shows a strong peak at 7 HHz with a weaker harmonic at 13 HHz as expected from increased derivative width. Figure 10 (left) is the auto correlation of the same data. . It is essentially a reproduction of Figure 8 (right) and strongly confirms the signal. If a periodic, single frequency, signal with an SNR of at least 1 is passed through an autocorrelation it will be recognized and essentially reproduced with an index starting at 0.5 and oscillating between 0.5 and -0.5. A SNR of 3 to 4 generates an autocorrelation response similar to that in Figure 10. A noise-free signal will be normalized and vary between +1 and -1. Pure noise results in the index randomly varying rapidly between 0.1 and -0.1. By maximizing the autocorrelation we optimized the SNR for the 7 HHz component. This required a 32-bin derivative followed by 0.06 smoothing. We evaluate, directly from the data, an average SNR of approximately 3 for our scale factor oscillation between $t = 0.55 - 0.80$.

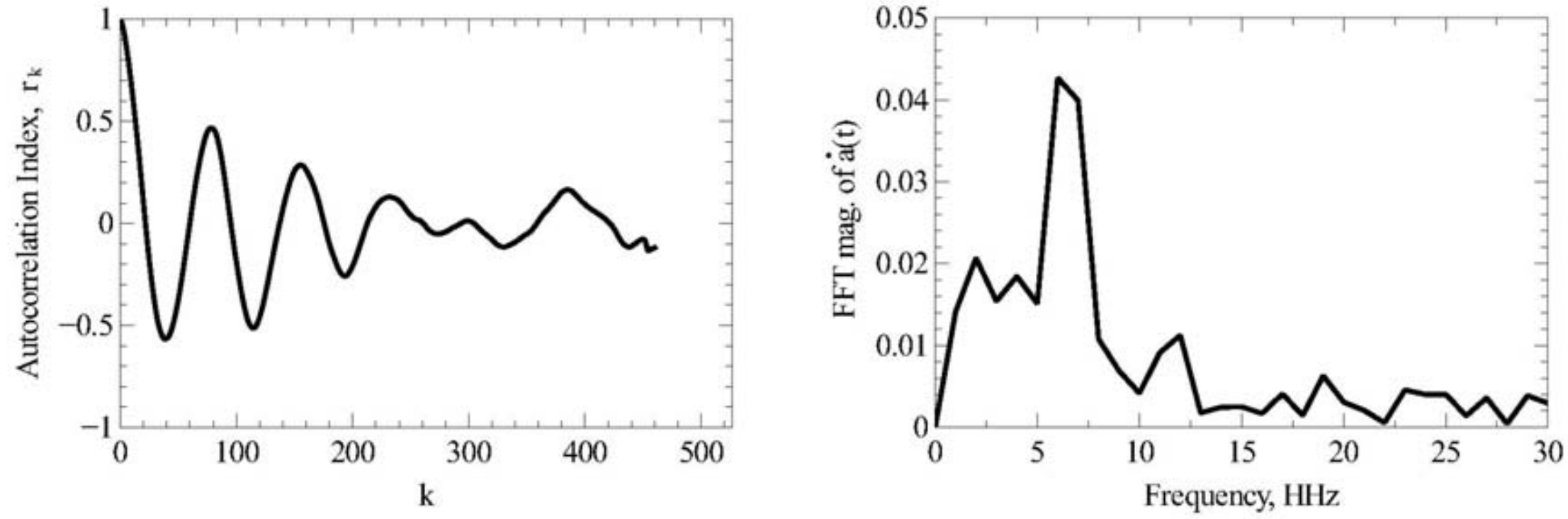

**Fig. 10.** Autocorrelation of scale factor derivative data, $\dot{a}$, (left) and discrete Fourier transform $f(\dot{a})$ (right). The autocorrelation k-axis corresponds to lookback times 0.4 ($k = 0$) to 1($k = 527$).

In Figure 11 we show a final comparison of the observed 7 HHz signal $\dot{a}(t)$ amplitude to the scalar field model $\dot{a}(t)$ amplitude from Section 5 with error estimates. The present model fits frequency quite well but inadequately describes amplitude. Also, the noticeable deviation in the oscillation approaching $t = 0.4$ is due to the significantly higher noise arising from the redshift sparseness.

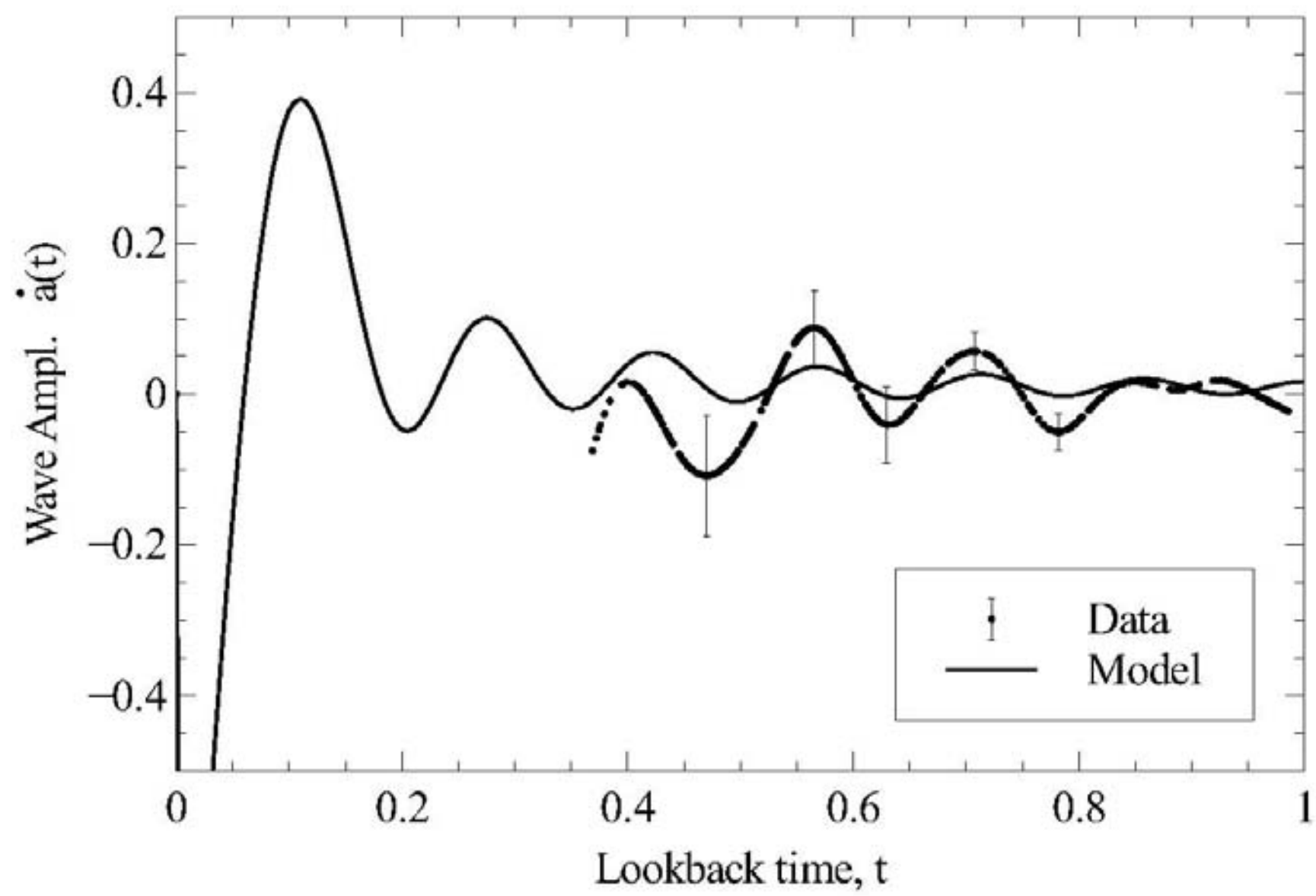

**Fig. 11.** Comparison of scalar field model of $\dot{a}(t)$ with observed 7HHz data. The model signal seen here is the full scalar $\dot{a}(t)$ signal of Fig. 12 with ΛCDM subtracted.

## 5. A COSMOLOGICAL SCALAR FIELD MODEL OF DARK MATTER

Ever since Guth introduced inflationary cosmology in 1981 (Guth, 1981), relating it to postulated scalar fields describing phase transitions, other authors ( Linde, 1982, 1983, 2014; Bardeen, Steinhardt & Turner, 1983; Steinhardt, Wang & Zlatev, 1998; Khoury, et al. 2001) advancing his proposal toward "new inflation", "chaotic inflation" and entirely new concepts have developed various scalar field cosmological models – all focused on the early inflationary transitions. A few authors considered the possibility of inflaton-like fields permeating the recent epoch. The first papers addressing this possibility appear to have been by Peebles & Ratra (Peebles & Ratra, 1988; Ratra & Peebles, 1988). They focused on utilizing scalar fields to represent essentially today's dark energy. Their work (Ratra & Peebles, 1988) foreshadows the present model. It recognizes that the scalar field particle mass at recent cosmological scales must be exceedingly small and even presents a calculation predicting a mass of the order of $10^{-32}$ eV in the present epoch – consistent with the scale of our model.

We basically combine Linde's chaotic inflation, using a simple quadratic harmonic oscillator potential, with Peebles' approach - moving the scalar field, $\phi(t)$, into the present era, but applying it to dark matter instead of dark energy. The dark energy then appears simply as a constant density. This, however, is not the first paper referring to scalar field dark matter. Matos, together with many co-authors, has pressed this concept for some time (Suárez, et al., 2014). Their work describes "fluctuations" in the scalar dark matter field as a function of the scale factor, $a(z)$. They do not suggest the possibility that the scale factor itself oscillates – as it should through the coupling with the scalar field. Though their scalar field mass is of order $10^{-24}$ eV they show that their scalar field replicates ΛCDM at early times and thus is capable of describing dark matter.

We describe oscillations in time directly with our model and match the frequency (mass of the scalar field) with the observed data. Our model is tightly constrained and effectively only one parameter, the frequency, controls the matching fit. We demand $a(1)=1$ and $\dot{a}(1)=1$ by definition. The choice of the initial scalar field amplitude, $\phi(0)$ fixes both of these conditions simultaneously once the appropriate density parameters are inserted. We show (Appendix B) that in the coupled equations $\dot{\phi}(0)=0$, leaving only the frequency to match the observed oscillations. There is apparently no phase adjustment. The frequency either matches or does not. We do in fact achieve good phase matching at the selected frequency. However, we have no control over amplitude once $\phi(0)$ is adjusted to set the scale factor conditions. We find that our observed amplitude is approximately twice the model value and we do not attempt to derive the harmonic signals – so the overall fit is incomplete. Further evaluation is outside the scope of this presentation.

### 5.1 Scalar Field Model

Is there a simple model which can account for the oscillations in $a(t)$ revealed in the last few sections? Since we are looking for an oscillating solution, it makes sense to couple the Einstein equations with an oscillating scalar field $\phi$ whose energy content can drive the evolution of $a(t)$. This kind of coupling is routine in models of inflation (Linde, 2014) where one is interested in the earliest evolution times with consequent density fluctuations. Here we will take the model seriously for all post-inflation times including the recent epoch (Ratra & Peebles, 1988). Our Lagrangian is chosen to be that of a simple harmonic oscillator,

$$\mathcal{L}=a^3\left(\frac{1}{2}\dot{\phi}^2-\frac{1}{2}m^2\phi^2\right). \tag{4}$$

We will see that this simple model contains the main features we wish to include. We follow well-known standard treatments ( Ratra & Peebles, 1988; Linde, 2014; Coles & Lucchin, 2002). Pressure and densities are,

$$P_\phi=\frac{1}{2}\dot{\phi}^2-\frac{1}{2}m^2\phi^2 \tag{5}$$

$$\rho_\phi=\frac{1}{2}\dot{\phi}^2+\frac{1}{2}m^2\phi^2 \tag{6}$$

The coupled scalar field-Friedmann equations are,

$$\dot{a}/a=\sqrt{\Omega_\Lambda+\Omega_b/a^3+\Omega_d/a^3+\Omega_r/a^4+\Omega_\phi}\ , \tag{7}$$

$$\ddot{\phi}+3\left(\frac{\dot{a}}{a}\right)\dot{\phi}+m^2\phi=0\ . \tag{8}$$

In eqn. (7), $\Omega_\phi = \rho_\phi / \rho_c$. All other $\Omega s$ are also scaled by the critical density for flat space, $\rho_c$. The subscript "$\Lambda$" refers to dark energy, "b" is baryonic matter, "d" is dark matter and "r" is radiation (including neutrinos). In the above, $a(t)$ is scaled such that $a(1)=1$ and $\dot{a}(1)=1$. Since either Planck or $\Lambda$CDM data suggests that space is flat, we must guarantee somehow that $\Omega = \sum_i \Omega_i = 1$ at $t=1$ (one Hubble time) as implied by (7). *In all that follows we set* $\Omega_d = 0$, that is, we omit dark matter. Our $\phi$-field energy , $\Omega_\phi$ (from (6)), will take its place. Otherwise, we adopt the Planck values of the constants: $\Omega_\Lambda = 0.683$, $\Omega_b = 0.047$, $\Omega_r = 0.00007$. The value of $\Omega_\phi(1)$ (the energy in $\phi$ now) will be determined by solving the coupled differential equations. What must be reproduced by these remaining constants are the values of $a(1) = \dot{a}(1) = 1$ and the observed frequency of oscillations in $a(t)$.

We take as the initial condition for the cosmic evolution factor to be $a(0)=0$. To be determined by best fitting to the data are the remaining constants of *m* (angular frequency), $\phi(0)$ and $\dot{\phi}(0)$. However, detailed, small-time dominance analysis of the coupled equations (7,8) shows that the only value of $\dot{\phi}$ consistent with the value $a(0)=0$ is $\dot{\phi}(0)=0$ (Appendix B). In view of these considerations, there remain two free parameters to fix: *m* and $\phi(0)$.

## 6. RESULTS OF THE NUMERICAL MODEL

After numerically solving (Endnote) equations (7, 8) we find $\phi(0) = 1.092$ and $m = 21.84$ as close to the optimal values required by the foregoing conditions. The value of $\Omega_\phi(1) = 0.256$ is very close to the Planck value of the dark matter density. There are now no free parameters.

In Figure 12 we show a plot of $a(t)$ and $\dot{a}(t)$ versus the lookback time. As we show in a moment, the fit of $a(t)$ vs *t* is excellent over the full range of data. It is difficult to see the oscillations in $a(t)$ on the scale of the plot, so the derivative $\dot{a}(t)$ is also shown clearly revealing the oscillations which have a frequency of twice that of the scalar field: $f_a = 2 f_\phi = 2(m/2\pi) = 6.95\, HHz$.

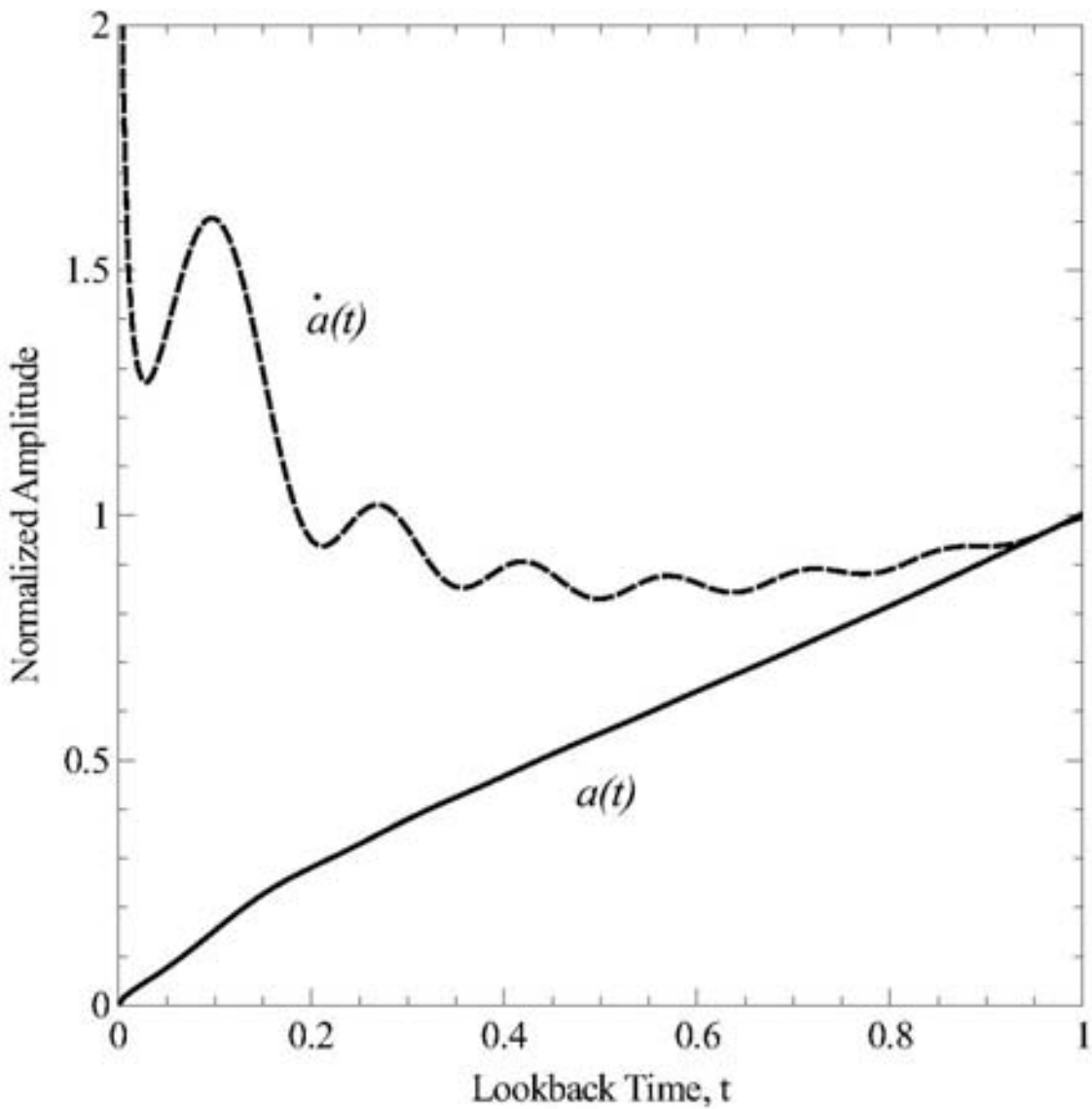

**Fig.12**. Plot of $a(t)$ and $\dot{a}(t)$ against lookback time to display the oscillations.

The fit to the actual $a(t)$ versus lookback time against the full set of supernova data is shown in Figure 13. As can be seen the Scalar Field plot (solid red line) oscillates about the WMAP- $\Lambda$CDM model. Nonetheless, the agreement with the data is excellent. There is significant deviation between the two models only at the early times, $t \leq 0.2$ .

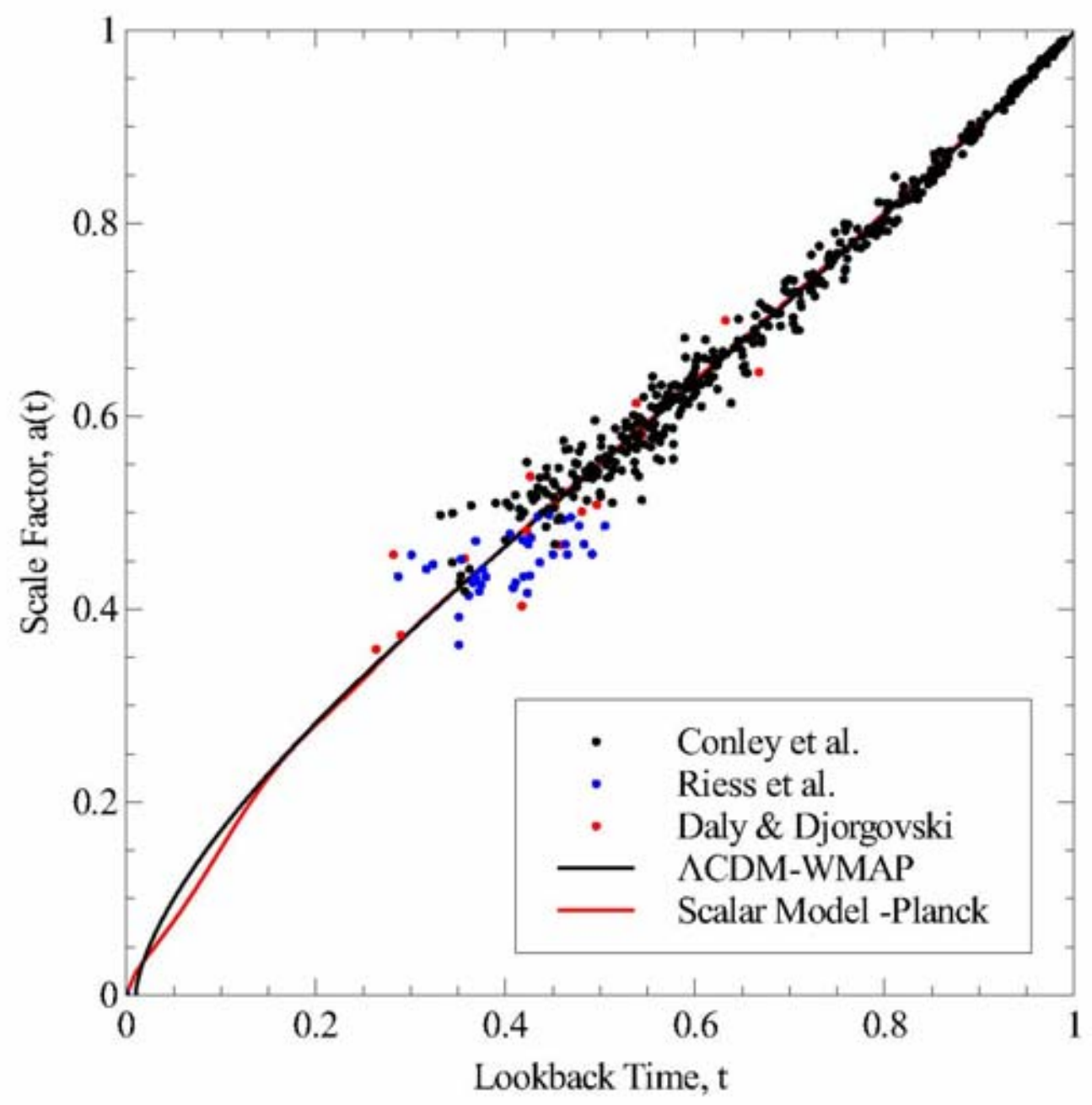

**Fig. 13.** Comparison of the $\Lambda$CDM a(t) model with the scalar field a(t) model together with the SNe data. R-squared goodness of fit is 0.98.

The scalar field and its energy oscillate at $f_\phi = m/2\pi = 3.48\ HHz$. We plot the energy in the scalar field in Figure 14 which has the Planck dark matter value of $\Omega_\phi = 0.256$ at the present time. The energy is quite large, but finite at $t = 0$. Superposed on this graph is that of a curve with $1/t^2$ falloff. It was pointed out in a previous publication (Ringermacher & Mead, 2014) that a satisfactory evolutionary model of the universe could be constructed by replacing dark matter by an energy density which fell off as $1/t^2$. This "toy" model virtually duplicates the standard $\Lambda$CDM model. Our current scalar field model seems to have this asymptotic feature; the deviation from this behavior occurs only from times less than $t \approx 0.2$.

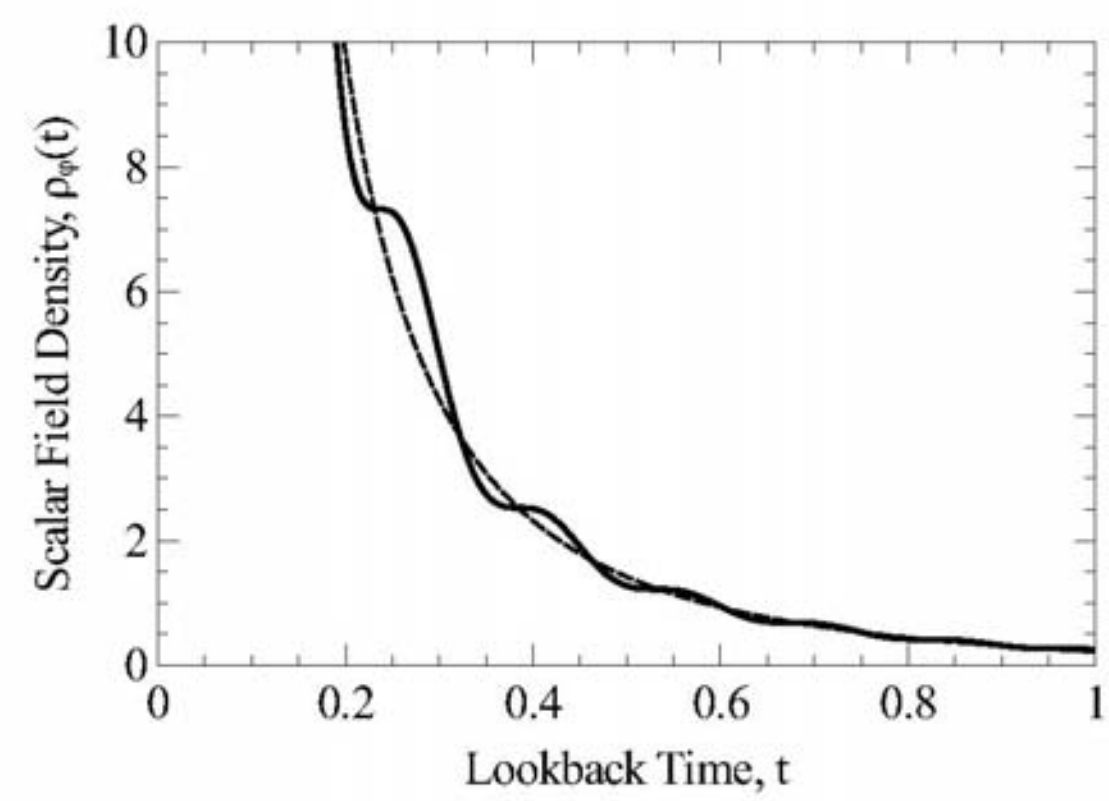

**Fig. 14**. Energy density of the scalar field (wiggly line) displaying a $1/t^2$ dependence on average. $\rho_\phi(0) = 284.4$

Related to the $\phi$-field energy is its equation of state parameter, $w_\phi$, defined as the ratio of the pressure of the dark matter field to its energy density $w_\phi = P_\phi(t)/\rho_\phi(t)$. Figure 15 shows this parameter as a function of lookback time.

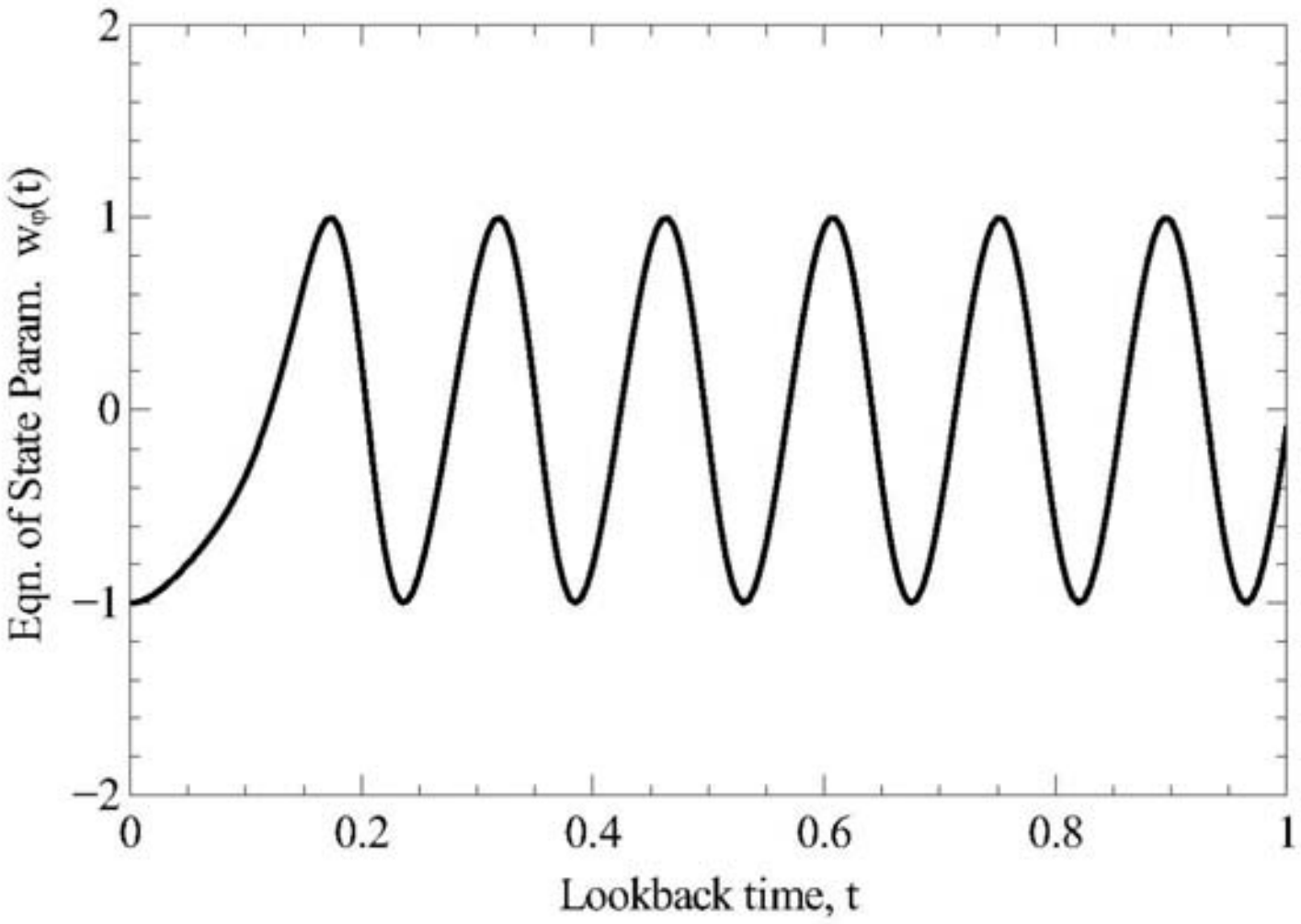

**Fig.15.** Equation of state parameter for the scalar field vs. lookback time.

The initial value of $w_\phi$ is exactly −1; the $\phi$-field energy acts initially as dark energy, enhancing the early expansion. After an initial increase, $w_\phi$ oscillates between −1, where the rate of growth of the universe is enhanced, and +1 where the rate of growth is slowed; there the field acts as a kind of "anti-dark energy". The mean is $w_\phi = 0$ which is zero pressure "dust" as in standard cosmology. The current value is $w_\phi(1) = -0.10$.

Also of interest to cosmologists is $q(t)$ the deceleration parameter defined from,

$$q(t) = -\frac{a(t)\,\ddot{a}(t)}{\dot{a}(t)^2}. \qquad (9)$$

A plot of $q(t)$ versus $t$ is shown in Figure 16. $q$ begins at the value +1, increases then oscillates rapidly with a decaying amplitude which is asymptotic to $q = -1$; it becomes virtually that value at around three Hubble times indicating a continued exponential growth of the universe. More important is the predicted value at the current time of $t = 1$ which is $q(1) = -0.578$ which is very nearly the accepted value with Planck parameters.

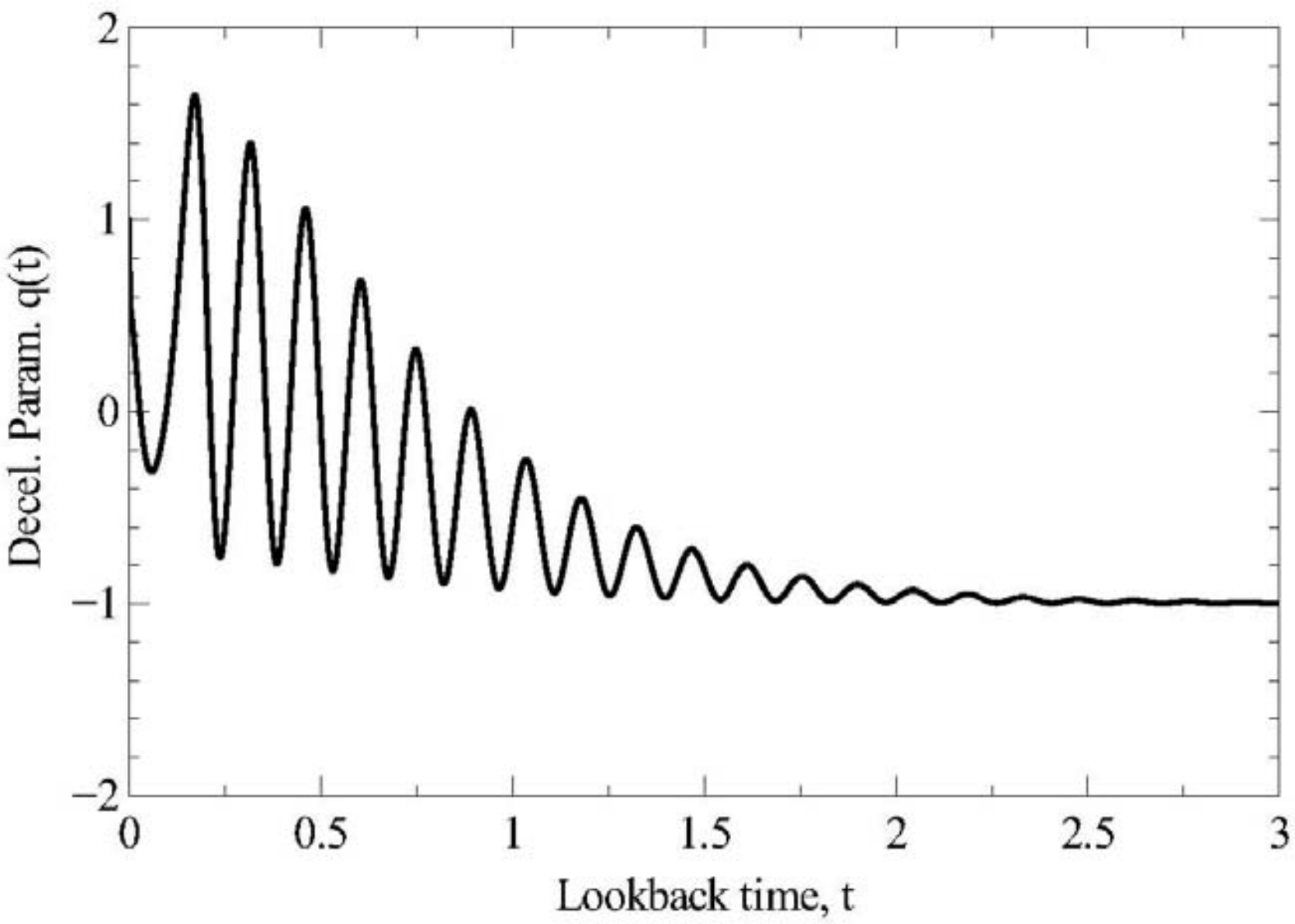

**Fig.16**. The deceleration parameter as a function of lookback time.

## 7. CONCLUSIONS

We have analyzed a model-independent plot of scale factor vs. lookback time for standard candle data and found discrete, damped, oscillations at 6.5 ± 0.5 HHz ( ~7 cycles in 13.8 Gy) and the second and third harmonics at 13 and 20 HHz. As was described, the standard candle data was compiled from three sets. The signals appeared mostly in the Conley set. Thus the data junction was not the source of the signal. The data were analyzed using four methods; smoothing, Fourier analysis, statistical noise analysis, and autocorrelation. Smoothing alone revealed the 7 HHz signal but such an analysis is not definitive. The Fourier analysis revealed the same 7 HHz signal along with its second and third harmonics prominently. Furthermore, the unequally spaced data were binned in equal-time spacings for the FFT analysis. This eliminates unequal spacing as the cause of the signals. Statistical analysis revealed a 10% probability that any single frequency within the smoothing bandwidth could be generated from normal random noise alone. We could not reproduce the 7 HHz signal together with harmonics as an artifact of any particular analysis. Thus it is unlikely that the three signals could be simultaneously generated by noise alone. Finally an autocorrelation was performed on the $\dot{a}(t)$ data. This showed a very strong response consistent with the insertion of a real signal as tested by trials with varying SNR. Inserting simulated random noise data always produced a noise output. So this is a very powerful test. It remains to reproduce the present results using an independent SNe data set.

We also constructed a simple harmonic oscillator scalar field model to account for the oscillations in the scale factor into the present epoch. The resultant oscillations matched the observations very well, even though the model was tightly constrained with no phase adjustment – only frequency. The amplitude adjustment was fixed by the requirement of normalization of $a(t)$ and $\dot{a}(t)$. Since the scalar field was coupled to the Friedmann equations, we showed that the oscillations followed, nearly exactly, the $\Lambda$CDM $a(t)$ curve to the present time while generating all the Planck $\Omega s$. Moreover, we found that $\Omega_\phi(1) = 0.256$ and that the deceleration parameter is $q(1) = -0.578$, both values of which are close to the $\Lambda$CDM values. This was accomplished by substituting $\Omega_\phi$ in place of $\Omega_d$ in the model. Thus we conclude that the scalar field manifests itself as the dark matter.

If the scalar field is a quantum field, then the 7 HHz frequency and its cosmological wavelength correspond to a scalar field particle mass of $3x10^{-32}$ eV, a number mentioned by previous authors. The model leaves open some questions. It does not account for the harmonics, though an extension to coupled oscillators is possible. The amplitude produced is also a factor of three too small. Further work on the model is out of the present scope of this paper at this time.

It was mentioned in Section 6 that our model deviates from $\Lambda$CDM only in early times, $t < 0.2$. At $t = 0.05$ an intense deceleration occurred (Fig.16) that caused the expansion rate (Fig.12) to fall well below the standard model. This could account for the observed surprisingly early galaxy formation that followed. A test of the present observations might be the appearance of these time-periodic components in the large-scale structure of the universe.

## ACKNOWLEDGEMENTS

We wish to thank Michael Vera for suggesting the use of Fourier analysis. We also wish to thank Judith Keating Ringermacher for her assistance in programming the data binning in Excel™.

## REFERENCES

Endnote: All computations have been verified with Maple™, Mathematica™ and Mathcad™ .

## APPENDIX A - Wide Baseline Differentiation

We define a 3 point numerical derivative of a function $f(t_i)$ centered at point $t_i$ as :

$$f_i'(t_i) = \frac{f_{i+1} - f_{i-1}}{t_{i+1} - t_{i-1}} = \frac{f_{i+1} - f_{i-1}}{2\Delta t} \tag{A1}$$

This derivative spans 2 time bins. We define a wide-baseline 3 point derivative similarly:

$$F_i'(t_i) = \frac{f_{i+n/2} - f_{i-n/2}}{t_{i+n/2} - t_{i-n/2}} = \frac{f_{i+n/2} - f_{i-n/2}}{n\Delta t} \tag{A2}$$

This derivative spans n time bins where n is an even number. Assume the function consists of signal plus noise, $f = S + N$ . Then,

$$F_i'(t_i) = \left( \frac{S_{i+n/2} - S_{i-n/2}}{n\Delta t} \right) + \left( \frac{N_{i+n/2} - N_{i-n/2}}{n\Delta t} \right) \tag{A3}$$

It is easy to show that the result obtained from taking a wide-baseline 3-point derivative of a function (1st term of (A3)) is not significantly different in amplitude from taking a normal centered 3-point derivative until the baseline gets extremely wide. The difference lies in the noise derivative (2nd term of (A3)). Thus the first term of (A3) is essentially the same as the signal derivative component of (A1). But the noise differential is the same for a 2-bin derivative or an n-bin derivative since it is merely the random noise summed at two points. So the numerator of the second term of (A3) is identical to the noise numerator of (A1). But the noise derivative denominator is now n/2 greater for (A3). This results in an n/2 noise reduction. Thus wide baseline differentiation acts as a noise filter with dramatic noise reduction. Figure A1 shows an example of a 12-bin derivative applied to a synthesized function with peak noise approximately equal to its amplitude (Fig. A1, left). The 2-bin derivative and 12-bin derivative have the same amplitude scales. The 12-bin derivative is, in effect, like taking a derivative of a noisy signal while not increasing its noise.

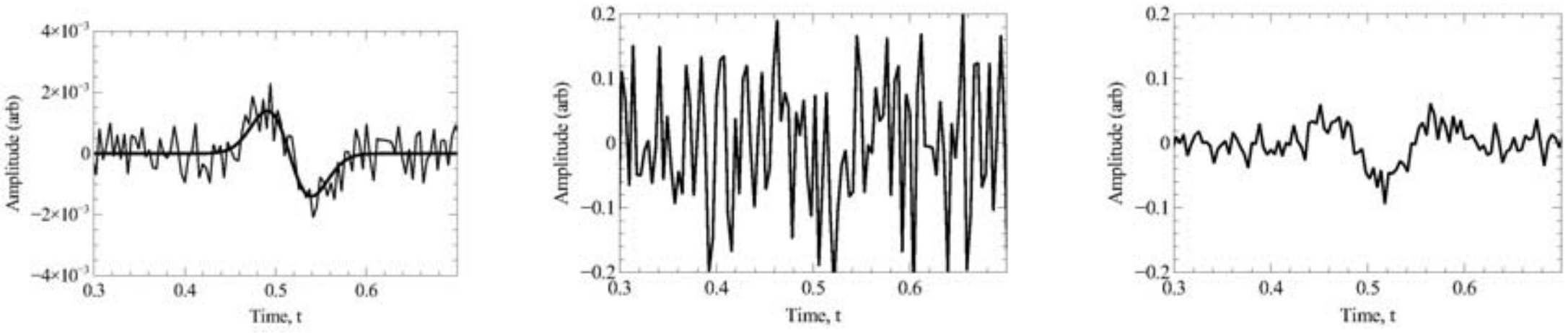

Fig.A1. Signal plus noise, left; 2-bin derivative, center;12-bin derivative same scale right.

## APPENDIX B - Solution of the model for short times

Equations (7,8) may be approximately solved for sufficiently short times. In (7), since $a \to 0$ at $t = 0$, the radiation term should dominate for short times even in the presence of the $\phi$-energy term which remains finite at $t = 0$. Thus, $\dot{a}/a \propto 1/a^2$ whose solution is $a(t) = A\,t^{1/2}$ for some constant, *A*. Equation (8) for short times then reads

$$\ddot{\phi} + \frac{3}{2t}\dot{\phi} + m^2\phi = 0 \qquad \text{(B1)}$$

This is a form of Bessel's equation with regular solution (Gradsteyn & Ryzhik)

$$\phi(t) = B\frac{J_{1/4}(mt)}{t^{1/4}} \qquad \text{(B2)}$$

for some constant B. The derivative of this function vanishes at t = 0; hence our initial condition that $\dot{\phi}(0) = 0$.

## ERRATUM [AJ, 149, 137 (2015), arXiv: 1502.06140 ]

**All references in the paper to "lookback time" are mislabeled and should be replaced by "cosmological time" including all graphs.**